\documentclass[12pt]{iopart}

\usepackage[pdftex]{graphicx}
\begin{document}

\title[LHC Nobel Symposium Summary]{Summary of the Nobel Symposium on LHC Results}

\author{John Ellis}

\address{Physics Department, King's College London, Strand, London WC2R 2LS, UK; \\
Theory Division, CERN, CH 122 Geneva 23, Switzerland}
\ead{John.Ellis@cern.ch}
\begin{abstract}
This is a personal summary of points made during, and arising from the symposium,
drawing largely from the talks presented there. The Standard Model
is doing fine, including QCD, the electroweak sector and flavour physics. The good news is that a
Standard-Model-like Higgs boson has appeared with the predicted mass, but the bad news is that  the
LHC has provided no hint of other new physics. On the other hand, physics beyond the
Standard Model is needed to explain neutrino masses and mixing, and we look forward to experiments
revealing their mass hierarchy and CP violation. Cosmology and astrophysics also require
new physics to explain the origin of matter, dark matter, dark energy and the CMB
fluctuations. I still think that supersymmetry is the best-motivated extension of the
Standard Model, and look forward to the next phases of LHC operation at higher energy 
and luminosity. In the mean time, ideas abound for a new accelerator to study the
Higgs boson in detail and/or whatever else the LHC may reveal in its next run.
~\\
~\\
~\\
\begin{centering}
KCL-PH-TH/2013-30, LCTS/2013-21, CERN-PH-TH/2013-218
\end{centering}
\end{abstract}

\maketitle

\section{Introduction}
The LHC experimental programme is aimed primarily at fundamental questions in
particle physics beyond the Standard Model: What is the origin of particle masses,
and is the answer manifested in a Higgs boson?
How do matter and antimatter differ, and is this related to
the reason why there are so many types of matter particles?
What is the dark matter in the Universe?
How are the fundamental forces unified?
What is the underlying quantum theory of gravity?

It is a remarkable success that already, in the first couple of years of operation of the
LHC at reduced energies, the answer to the first of these questions has been provided,
and there are reasons to think that its answer may be linked to answers to some of the
other fundamental questions.
Now, during the first long shutdown of the LHC before it restarts at high energy, is a
good moment to review what we have learnt so far, and discuss the prospects for
further advances in coming LHC runs and at possible future colliders.

In this talk, I start by reviewing the successes of the Standard Model, in particular
the triumph of the discovery of a new particle that resembles closely the Higgs boson
of the Standard Model, before discussing the need for new physics to explain
neutrino masses and mixing as well as many cosmological puzzles. I persist in
thinking that supersymmetry is the best-motivated extension of the Standard Model
that may be accessible in the future runs of the LHC. In the mean time, one should consider
ideas for future accelerators that could measure better than the LHC the properties of the 
Higgs boson, as well as study whatever other new physics it may reveal.

\section{QCD}
This is basis for physics at the LHC~\cite{AdR}. In addition to its intrinsic interest and its
r\^ole in generating backgrounds via jets and pile-up, understanding QCD is essential for
calculating the production of new particles such as the Higgs boson and yields
possible signals for new physics such as boosted jets and events with rapidity gaps.
Moreover, a better understanding of multiparticle production in QCD would also be helpful
to our colleagues studying cosmic-ray physics~\cite{RO}.

Specific instances where understanding the underlying event are important include the
accurate measurement of $m_t$ and $M_W$, where good understanding of the parton
distributions in the proton are required, and problems of pile-up and colour reconnection arise.
Instances where boosted jets and jet substructure are important include the search for signatures of baryon-number-violating
$R$-parity violation and hadronic Higgs decays. 
In addition to being interesting in their own right, events with single and multiple rapidity gaps
may also be important for understanding diffraction and hence the total hadronic cross section, and
double-diffractive production of the Higgs boson would be an interesting tool for determining the CP
properties of the Higgs coupling to gluons. Better understanding of particle production in the
forward direction is interesting to those modelling high-energy cosmic-ray collisions~\cite{RO}.
Finally, we also note the appearance in $pp$ collisions
of a mysterious near-side ridge~\cite{AdR} that may cast light on the origin of the similar effect in heavy-ion
collisions~\cite{JS}.

The cross sections for the underlying hard-scattering processes are under relatively good
control, with many NNLO perturbative QCD calculations now available~\cite{TS}, and new input coming from string methodology~\cite{ER}.
The agreement with measurements of jet production is impressive, extending over many orders of
magnitude to jet transverse momenta $p_T > 2$~TeV~\cite{AdR}, and interesting measurements of the running of 
$\alpha_s$ at high energies are now possible. Precision is important for both the discovery and
interpretation of any new physics. For example, it would be good if the convergence of the perturbative expansion
for $gg \to H$ could be improved, with either higher-order calculations or a more cunning choice of
renormalization scheme. Also, more precise understanding of initial-state radiation (ISR) would be
invaluable for some new physics searches using monojets, e.g., dark matter and models with compressed spectra.

Progressing to larger distances, while progress has been made in the matching of matrix-element calculations with parton
showers, there are still significant differences between different codes, e.g., in calculations of the Higgs
$p_T$ distribution. Meanwhile, the hadronization process is still problematic: it was said here that there had been ``no new
ideas in 30 years" in the long-running debate between string and cluster models~\cite{TS}. Both these need to be understood
better if, e.g., the full potential of boosted-jet techniques is to be realized. Which experimental measurements would best
help the experts find better, universal tunes of their Monte Carlo generators?

Relativistic heavy-ion collisions at the LHC are revealing interesting new collective phenomena in hot
and dense nuclear matter~\cite{JS}. They offer the possibility of exploring QCD thermodynamics via
phenomena such as the Quark-Gluon Plasma (QGP), chiral symmetry breaking (CSB), and
deconfinement. Particle abundances are generally in good agreement with simple thermodynamic
models with a freeze-out temperature $T \sim 164$~MeV similar to that measured in lower-energy
collisions and a smaller baryon chemical potential $\mu_B \sim 1$~MeV. The production of strange 
particles is enhanced as expected, but there are fewer protons than expected in this model.
There is evidence also for direct thermal photon production inside the plasma before freeze-out, with a higher effective temperature than at
RHIC: $T^\gamma_{LHC}/T^\gamma_{RHIC} \sim 1.37$. The data also suggest the suppression of
$\Upsilon$ production as expected in the QGP, whereas the production of the $J/\psi$ provides evidence
for $c {\bar c}$ recombination in the QGP.

However, big surprises in relativistic heavy-ion collisions have been provided by bulk
measurements such as those of the elliptic flow $v_2$ and higher moments $v_n$ of the hadron azimuthal distribution,
which suggest that the QGP has very small shear viscosity $\eta$, and by measurements of jet quenching~\cite{BM}. In retrospect,
the small value of $\eta$ in low-temperature quark-gluon matter should have been anticipated. At very low
temperatures nuclear matter can be described as a liquid, but at higher temperatures $\sim 100$~MeV it can
be considered as a meson gas. These are weakly-interacting at low temperatures, but interact more strongly at
higher temperatures because of chiral symmetry, leading to a calculable reduction in $\eta$ as the quark-hadron
transition temperature is approached from below, as seen in~\cite{sPHENIX}. 
Likewise, at very high temperatures the the QGP is expected to 
behave like an asymptotically-free gas that becomes more strongly-interacting at lower temperatures close to
the quark-hadron transition temperature, again corresponding to a low value of $\eta$, as also seen in~\cite{sPHENIX}. 
The surprise was that
the viscosity $\eta$ would be so low in the transition region, close to the lower bound $\eta \ge 1/4 \pi$ derived on the basis of the
AdS/CFT correspondence~\cite{ER}, corresponding to a liquid that may be more perfect than any other known.

\begin{figure}[h!]
\centering
\hspace{-1cm}
\includegraphics[scale=0.4]{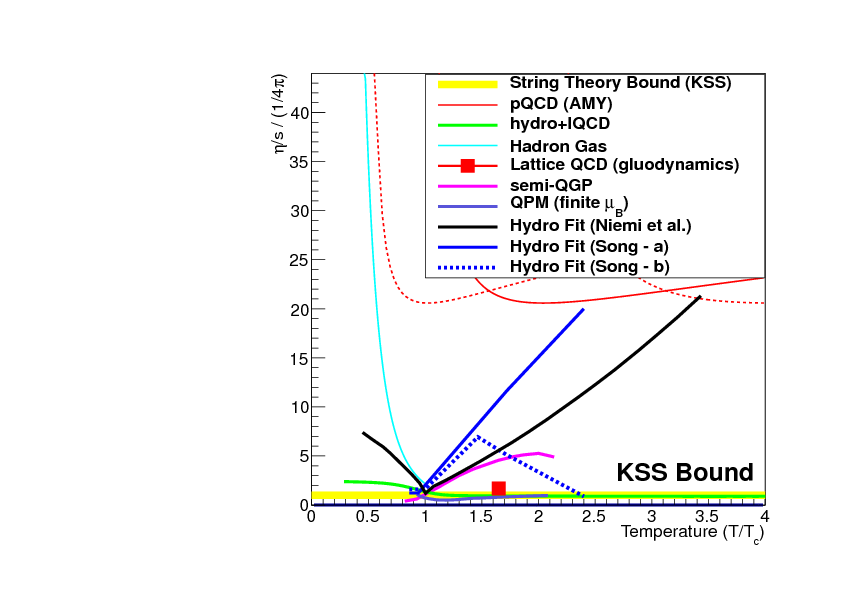}
\caption{\it The ratio of the shear viscosity divided by the entropy density, $\eta/s$, of quark-gluon matter,
relative to the AdS/CFT bound (denoted by KSS) around the critical temperature, $T_c$,
according to calculations in the hadron gas, on the lattice, in perturbative QCD and in more phenomenological
motivated by hydrodynamics. Plot taken from~\cite{sPHENIX}, where more details can be found.}
\label{fig:sPHENIX}
\end{figure}

One can regard relativistic heavy-ion collisions as providing a ``multiverse" of ``Little Bangs" with different
initial conditions that are measurable via the azimuthal moments $v_n$~\cite{JS,BM}, much as measurements of the
cosmic microwave background anisotropies $C_l$ give us insight into the primordial fluctuations
(generated by inflation?~\cite{RO,HM}). These measurements tell us that the low-temperature QGP is well described by
the hydrodynamics of an almost-perfect liquid with $\eta <  (2~{\rm to}~3)/4 \pi$. The picture of a strongly-interacting
liquid is reinforced by measurements of jet quenching and the dijet energy asymmetry. These tell us that the
medium is very effective at absorbing hadronic energy, but the fragmentation of the residual jets that punch through the medium
resembles strongly the fragmentation of lower-energy jets that fragment in vacuo.

\section{Electroweak Measurements}

Impressive progress has already been made at the LHC~\cite{KE}. For example, the LHC measurement of $m_t$
already has an uncertainty comparable to that at the Tevatron, namely $\sim 0.5$\%~\cite{RC}, and the greater statistics at the LHC offer
prospects for a substantial reduction, if the systematic errors can be mastered. 
This greater accuracy is certainly desirable, as we see shortly~\cite{RB,GP,GA}.
In addition to the dominant ${\bar t} t$ production process, where spin correlation measurements cast
light on the production process, single-top production has been measured, reassuring us
that $V_{tb} \sim 1$. We recall that the forward-backward production asymmetry $A_{FB}$
measured at the Tevatron is still problematic, and there are hopes that light will be cast by
measurements of the quantities $A_\ell$ and $A_c$~\cite{GP}. Meanwhile, searches
are on for associated $t + W$ and $t + H$ production. 

There are challenges in the measurement of $m_W$~\cite{KE} associated with pile-up (it was even suggested that a special low-pile-up
run be scheduled), uncertainties in the parton distribution functions (PDFs) at low $x$, the effects
of material in the detector, etc.. It will be a while before the Tevatron uncertainty in $m_W$ will be bettered.
Meanwhile, the LHC has provided its first measurement of $\sin^2 \theta_W = 0.2297 \pm 0.0010$: a long way
to go before the LEP uncertainty can be matched, but the LHC is still in its early days.

The LHC has already made several measurements of diboson production~\cite{KE}, and is starting to provide
competitive constraints on multi-boson couplings. However, improvements in the accuracy of
theoretical predictions will be needed before these constraints can be exploited fully. These
measurements can be used provide constraints on higher-order terms in the effective electroweak
Lagrangian that may be combined those provided by measurements of Higgs production and
decays. This is an example how
electroweak and Higgs coupling measurements complement each other in searches for New Physics.
So far there is limited sensitivity to triboson production and hence quartic gauge-boson
interactions, though an interesting constraint on the $\gamma \gamma W^+ W^-$ vertex is
provided by the new LHC measurement of $\gamma \gamma \to W^+ W^-$.

\section{Electroweak Theory}

Experiments at LEP already posed a paradox that has been accentuated by
experiments at the LHC~\cite{RB,GA,NAH,HM}: in the words of Enrico Fermi: ``Where is everybody?"
In other words: ``Why is there just a light (and apparently elementary) Higgs boson and
nothing else, specifically none of the plethora of new particles predicted in popular
theoretical extensions of the Standard Model?"
If there is some other low-mass physics: Why have we seen no indirect evidence of it?
Alternatively, if there is nothing light, is a light, elementary Higgs boson unnatural?

The discovery of a light, apparently elementary Higgs boson at the LHC has
accentuated the naturalness problem that has exercised theoretical minds for over 30 years.
The full particle content of the Standard Model is now established, and we know that the
quadratic divergences in loop diagrams are not cancelled, signalling in many minds
extreme sensitivity to the parameters of high-scale physics and a need for unreasonable (?)
fine-tuning.

That said, the experimental values of $m_H$ and $m_t$ have an intriguing property~\cite{RB,GP,GA}:
they lie very close to the boundary of the region where the effective potential of the
Standard Model is stable, probably in a region where our present electroweak vacuum is
metastable unless there is new physics at some scale below about $10^{10}$~GeV. Is
this a coincidence, or is it telling us something profound?

\section{Flavour Physics}

This is one example of an area where new physics has failed to manifest itself even
indirectly. The Cabibbo-Kobayashi-Maskawa (CKM) paradigm continues to work
very well~\cite{YN}, and dominates over any new physics contributions to the 15 non-zero modes
of CP violation that have now been measured~\cite{VG,PK}. There were hopes for a while that there
might be CP violation beyond the Standard Model in the $D^0$ system, but this hope
was always subject to theoretical uncertainties, and is not supported by more recent
LHCb results. 

Moreover, rare decays also agree with the Standard Model predictions,
the most recent being $B_s \to \mu^+ \mu^-$, which has now been observed 
by both LHCb and CMS~\cite{LHCbmumu,CMSmumu}. The 
one current instance of an apparent discrepancy with the Standard Model
is in one of the angular distributions in $B \to K^* \mu^+ \mu^-$ decay~\cite{Kstar}, but it is too
early to judge whether this will survive as evidence for new physics.

In the future, we will want to push the experimental
error down to the level of the theoretical error, below 10\%, and we look forward
to measurements of $B_d \to \mu^+ \mu^-$. Currently both CMS~\cite{CMSmumu} and LHCb~\cite{LHCbmumu} have
2-$\sigma$ hints for this decay, though at rates higher than expected in the 
Standard Model. It would also be interesting to measure $B_s \to \tau^+ \tau^-$.
LHCb is planning a major upgrade in 2018, aiming at measurements of CP
violation in $B \to \phi \pi$ and $D$ decays at the Standard Model level~\cite{VG}, and
CMS and ATLAS will continue their measurements of rare $B$ decays, while
Belle-II is on the horizon~\cite{PK}. This will complement the LHC experiments in many
channels involving photons and/or neutrinos, as well as in charm and $\tau$
decays. Thus, probes of the CKM paradigm will grow ever more
precise in the coming years. However, at the moment it seems that any new 
physics at the TeV
scale should copy the CKM model via some scenario with minimal flavour violation~\cite{YN}.

\section{The (NGA)EBHGHKMP Mechanism}

Following the introduction into particle physics of spontaneous global symmetry
breaking into particle physics by Nambu~\cite{N} and the formulation of a simple
field-theoretical model by Goldstone~\cite{G}, as well as the interpretation by Anderson~\cite{A} of superconductivity
in terms of  a spontaneously-broken local U(1) symmetry, in 1964 several papers
introduced spontaneously-broken local symmetry into particle physics. The initial paper
by Englert and Brout~\cite{EB} was followed a few weeks later by two papers written independently by Higgs: 
the first pointing out that a technical obstacle to a four-dimensional extension of Anderson's
approach could be circumvented~\cite{H1}, and the second proposing a specific four-dimensional model
with a massive scalar boson~\cite{H2}. The subsequent paper by Guralnik, Hagen and Kibble~\cite{GHK} referred
explicitly to these earlier papers. Also of note is a relatively-unknown 1965 paper by Migdal and Polyakov~\cite{MP},
which discusses the partial breaking of a local non-Abelian symmetry, ahead of the influential paper
of Kibble~\cite{K}. 

Of all these authors, Higgs was the only one who mentioned explicitly the existence of a
massive scalar boson (see equation (2b) of his second paper~\cite{H2}), and he went on to write a third
paper in 1966~\cite{H3} that discusses the properties of this `Higgs boson' in surprising detail including, e.g.,
its decays into massive vector bosons.

For a decade, very few people took seriously Higgs' prediction of this boson, there being only a
handful of papers about before the phenomenological profile that Mary Gaillard, Dimitri Nanopoulos
and I wrote in 1975~\cite{EGN}. Fortunately, the LHC community did not pay attention to the caveat in the last sentence
of our paper that ``we do not want to encourage big experimental searches for the Higgs boson", and
the ATLAS and CMS experiments announced the discovery of a candidate for a (the?) Higgs boson
on July 4th 2012~\cite{ATLAS,CMS}.

\section{From Discovery to Measurement}

The emphasis has now shifted from the discovery of this new particle to measurements
of its properties~\cite{AR}. Combined, their mass measurements yield $m_H = 125.6 \pm 0.3$~GeV,
somewhat below the cosmological stability bound, as already commented. The mass values
measured by ATLAS and CMS are quite consistent: there is some difference between the
values measured by ATLAS in the two high mass-resolution channels $\gamma \gamma$
and $Z Z^* \to 4 \ell^\pm$, but this is not significant.

\begin{figure}[h!]
\centering
\includegraphics[scale=0.5]{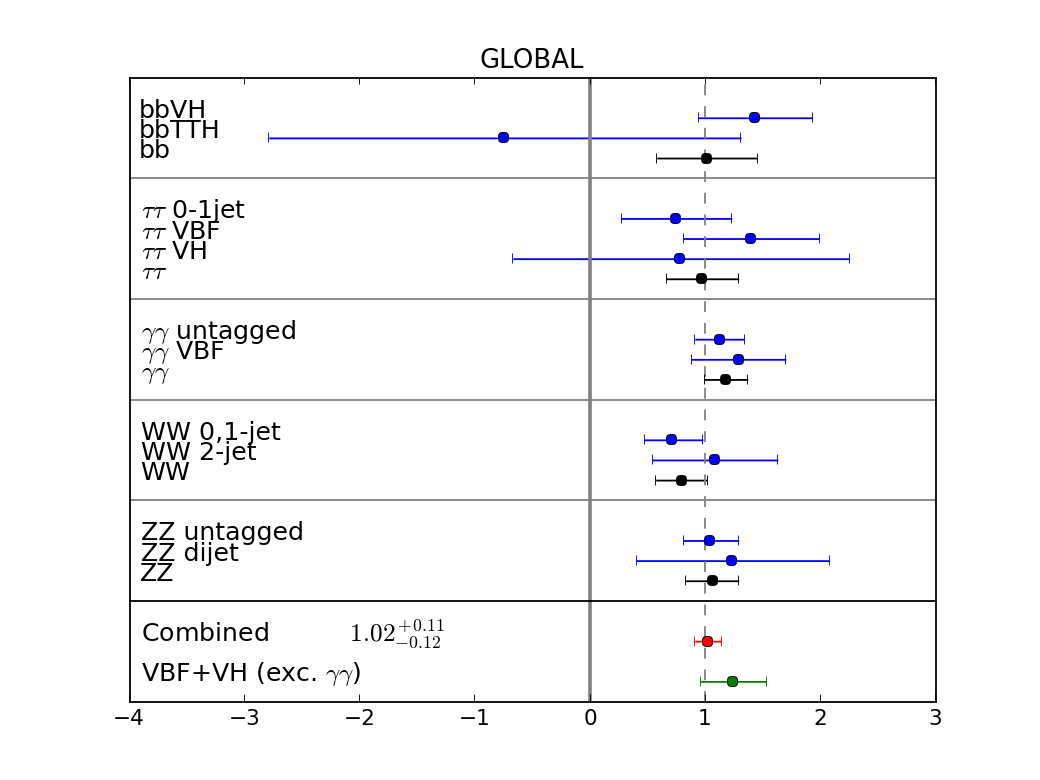}
\caption{\it A compilation of the Higgs signal strengths measured by the ATLAS, CDF, D0 and CMS
Collaborations in the ${\bar b} b$, $\tau^+ \tau^-$, $\gamma \gamma$, $WW^*$ and $ZZ^*$ final states~\cite{EY3}.
We display the combinations of the different channels for each final state, and also the 
combination of all these measurements, with the result for the VBF and VH channels (excluding the $\gamma\gamma$ final state) shown separately in the bottom line.}
\label{fig:EY}
\end{figure}

The signal strengths observed in many channels are compatible with the Standard Model
within the measurement errors, which are currently $\sim 20$ to 30\%, as seen in Fig.~\ref{fig:EY},
The mean signal strength, averaged over all the observed channels, is $\mu = 1.02^{+0.11}_{-0.12}$~\cite{EY3}.
Frontiers include production via vector-boson fusion (VBF), which has been observed with
significance $\sim 2 \sigma$ in several channels, and $\sim 3 \sigma$ combined.
The decays to $\tau \tau$ and ${\bar b}b$ are both emerging at the $\sim 3 \sigma$ level,
and there is indirect evidence that the $H{\bar t} t$ coupling has approximately the Standard Model
value, from the magnitudes of the $H gg$ and $H \gamma \gamma$ couplings. Direct evidence
for the $H{\bar t} t$ coupling will be sought via production of $H$ in association with ${\bar t} t$,
single $t$ or ${\bar t}$.

The situation now is reminiscent of a jigsaw puzzle that you have been working on for almost 50 years.
In your search for the last piece, you have found `behind the sofa' a crumpled piece of cardboard: 
does it have the right shape and size to be
the missing piece of the puzzle, or is it something different (which would be even more exciting)? If it {\it is}
the Higgs boson, how best to measure its properties with what sort of Higgs factory, and is it accompanied
by other (Higgs) particles~\cite{CM}?

\section{What is it?}

In our attempts to characterize the newly-discovered $H$ particle, we ask the following questions.
Does it have spin 0 or 2? Is it scalar or pseudoscalar? Is it elementary or composite? Does it couple to 
other particles proportionally to their masses? Are quantum (loop) corrections to its couplings in
agreement with the Standard Model? What are its self-couplings?

Various ways of distinguishing different spin hypotheses have been proposed, including
angular and kinematic distributions in $H \to \gamma \gamma, Z Z^*$ and $W W^*$ decays~\cite{AR},
the invariant mass distribution and energy dependence in associated $V + H$ production.
Simple graviton-like spin-2 couplings of the $H$ particle have been excluded in many ways.
Some of the same measurements also exclude the possibility that the $H$ particle has
predominantly pseudoscalar couplings, e.g., to $Z Z^*$. However, this possibility should be checked
in each channel, as models of CP violation may not yield universal combinations of scalar and
pseudoscalar couplings.

\begin{figure}[h!]
\centering
\includegraphics[scale=0.5]{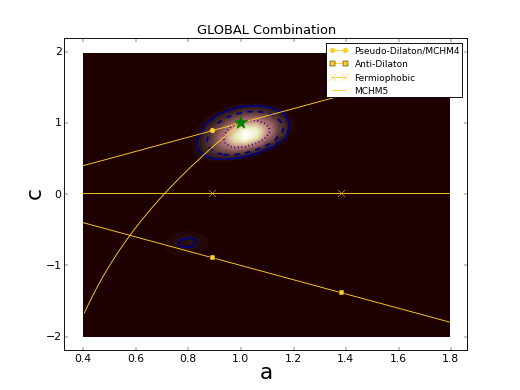}
\caption{\it The constraints in the $(a, c)$ plane imposed by a global combination of all the 
available Higgs coupling measurements in the ${\bar b} b$, $\tau^+ \tau^-$, $\gamma \gamma$,
$WW^*$ and  $ZZ^*$ final states~\cite{EY3}.}
\label{fig:EY2}
\end{figure}

Many theoretical groups as well as the experimental collaborations have confronted the data to models
in which the $H$ couplings to bosons and fermions are rescaled by factors $(a, c)$ relative to
the Higgs couplings in the Standard Model~\cite{GA}. Fig.~\ref{fig:EY2} shows the result of one such global
fit to all the available data on $H$ production and decays~\cite{EY3}. We see that there is no evidence for
any significant deviation from the Standard Model, and that alternative composite models
(represented by the yellow lines) are either excluded or constrained to resemble strongly the
prediction $(a, c) = (1, 1)$ of the Standard Model.

\begin{figure}[h!]
\centering
\includegraphics[scale=0.5]{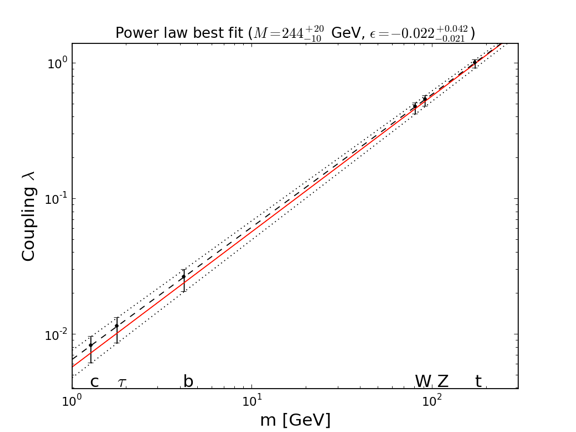}
\caption{\it The results of a global  fit using the two parameters $(M, \epsilon)$: the red line is the Standard Model prediction, the
black dashed line is the best fit, and the dotted lines are the 68\% CL ranges~\cite{EY3}.}
\label{fig:EY3}
\end{figure}

An alternative way to parameterize the $H$ couplings to other particles is
\begin{equation}
\lambda_f \; = \; \sqrt{2} \left(\frac{m_f}{M} \right)^{1 + \epsilon}, \; g_V \; = \; 2 \left(\frac{m_V}{M} \right)^{2 + 2 \epsilon} \, ,
\label{epsilonM}
\end{equation}
where in the Standard Model $\epsilon = 0, M = v = 246$~GeV. The result of a global fit using
the parameterization (\ref{epsilonM}) is shown in Fig.~\ref{fig:EY3}~\cite{EY3}. The Standard Model prediction
that the couplings should depend linearly on the particle masses is very compatible with the data:
the global fit yields
\begin{equation}
\epsilon \; = \; - 0.022^{+ 0.042}_{- 0.021}, \; M \; = \; 244^{+20}_{- 10}~{\rm GeV} \, .
\label{values}
\end{equation}
As we wrote in~\cite{EY2}, the $H$ particle walks and quacks like a Higgs boson.

At one stage there was some excitement that the loop couplings of $H$ to $gg$ and (particularly)
$\gamma \gamma$ might differ significantly from the Standard Model predictions. However, a
global fit currently finds consistency with the Standard Model at the 68\% CL,
though there is still some scope for the $H \gamma \gamma$ vertex to be larger than in the Standard
Model and for the $Hgg$ coupling to be somewhat smaller.

Almost all the items on the checklist in the first paragraph of this Section have now been verified, 
the only outstanding item on the checklist being the triple-boson coupling.
In view of the successful checks discussed in the previous paragraphs, beyond any reasonable doubt the
$H$ particle is a Higgs boson.  This is a Big Deal, and a tipping-point in particle physics. In order for a particle theory to
be calculable over many decades in energy or distance, it must be renormalizable.
We have known for over four decades that the most general renormalizable theory includes scalar bosons
as well as gauge vector bosons and fermions. Given the absence of a scalar boson and the technical
issues of naturalness that it raises, many theorists have doubted whether such a third element of the
renormalizable Holy Trinity actually exists. The discovery of a Higgs boson has caused considerable 
mortality among theories, if not among
theorists. Now we know that a/the Higgs boson does exist, we can focus on the next set of
phenomenological issues to be addressed by the LHC and possible future accelerators.

How closely does it resemble the Higgs boson of the Standard Model?
Are there other Higgs bosons~\cite{CM}? What are the upper limits on the couplings of any more massive Higgs bosons?
If this Higgs boson is not alone, is there an extra singlet boson beyond the Standard Model? Or is there a
second Higgs doublet? Perhaps there is a fermiophobic Higgs doublet? Perhaps a second Higgs doublet
appears as in the minimal supersymmetric extension of the Standard Model (MSSM)? Does this Higgs
boson have non-standard decays? Are there invisible decays? Does it decay into ${\bar b}b, \tau^+ \tau^-$
and $\mu^+ \mu^-$ as predicted in the Standard Model? Does it have flavour-changing decays, e.g., into
$\tau^\pm \mu^\mp$~\cite{BEI}? Is there any sign of an anomaly in $VV$ scattering?

Is the fact that $m_t$ and $m_H$ are measured to be very close to the stability bound
some sort of hint? (One might add that also the present value of the dark energy, a.k.a., the
cosmological constant, is very close to Weinberg's anthropic upper bound~\cite{W}.)
Supersymmetry is one example of new physics that could stabilize the electroweak vacuum
(and would also remove the most divergent contribution to the dark energy), in addition to its
other well-advertized benefits. But is the supersymmetric spectrum strongly split? Or is it
broken completely at a high scale? Should we modify or abandon the naturalness criterion?
Does Nature care about naturalness? But should we really be content if there is any one of
the myriads of vacua in the string landscape that describes our Universe, and accept it on
anthropic grounds, abandoning many of our scientific aspirations for deeper understanding? 

\section{Neutrino Physics}

This is another area where new theoretical ideas would be welcome.
There are plenty of models for neutrino masses and mixing~\cite{SP}, but none are particularly
convincing (not even mine). Dispiritingly, anarchy works just fine~\cite{HM}! As in the case of the
quark flavour problem, maybe we need yet more experimental clues.

For example, we do not yet know the answers to such basic questions as
whether neutrinos have Majorana or Dirac masses, nor whether they have a
normal or inverted mass hierarchy. The most tantalizing question is whether
there is CP violation in the neutrino sector. Both Majorana masses and CP violation
are important in principle for leptogenesis~\cite{HM}, although their discoveries would not be
sufficient to calculate it without additional input, either theoretical or experimental.

The good news is that neutrino experiments have prospects for addressing many
of these neutrino questions in the near future, as well as the possible existence of
sterile neutrinos and the reactor anomaly~\cite{JC}. Several experiments are ready to
determine the hierarchy of neutrino masses, and the relatively large value of
$\theta_{13}$ measured by the Daya Bay~\cite{DB} and RENO~\cite{RENO} experiments raises hope
that the  `Holy Grail' of CP violation may be within reach. As we heard here, an interesting
local possibility could be to exploit an intense 400-MeV neutrino beam from the
European Spallation Source in conjunction a large water {\v C}erenkov detector
located near here.

Another frontier in neutrino physics is in astrophysics. IceCube has recently
reported the observation of very-high-energy neutrinos that cannot have
just an atmospheric origin~\cite{IceCube}. Are these finally the harbingers of GZK physics,
or do they have some other source?

\section{Cosmological Inflation in Light of Planck}

\begin{figure}[ht]
\begin{centering}
\hspace{3cm}
\includegraphics[scale=0.6]{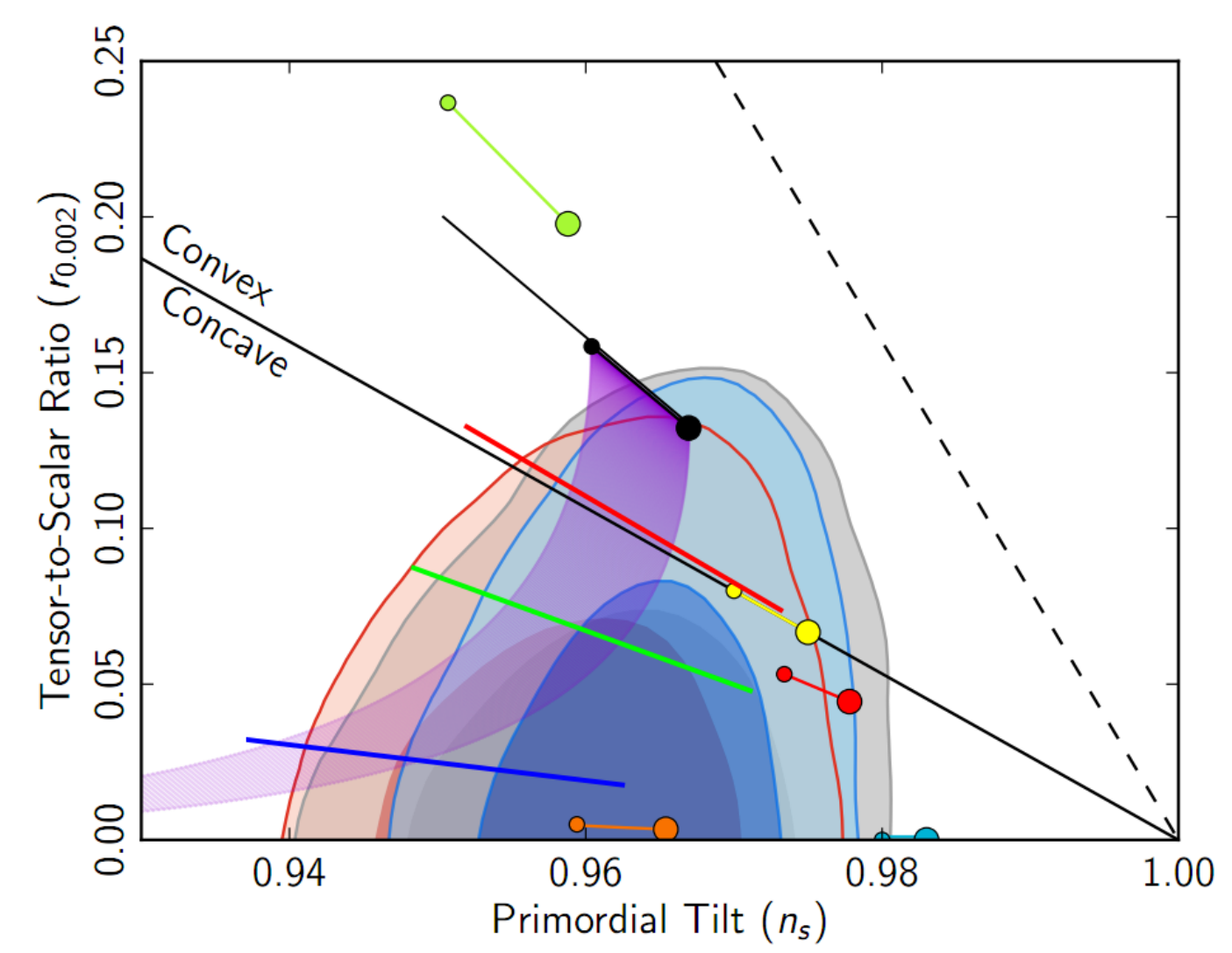} 
\end{centering}
\caption{\it
Predictions in the $(n_s, r)$ plane of various monomial
models for inflation with numbers of e-folds in the range $50 < N < 60$, taken from~\cite{Planck},
compared with a Wess-Zumino model for inflation
for various values of its parameters (red, green and blue lines) in the range $40 < N < 70$~\cite{CEM}.
Also shown are the predictions of the Starobinsky model~\cite{Starobinsky}: Higgs inflation~\cite{BS}
makes very similar predictions.}
\label{fig:inflation}
\end{figure}

The cosmic microwave background (CMB) is another window on ultra-high-energy
physics~\cite{RO,HM}. Its perturbations are thought to have originated in quantum fluctuations
during an early inflationary epoch driven by some scalar inflaton field $\phi$. The recent CMB 
observations with the Planck satellite~\cite{Planck} are consistent with inflation, in particular
because they confirm the tilt in the scalar perturbation spectrum:
\begin{equation}
n_s \; = \; 0.9603 \pm 0.073
\label{ns}
\end{equation}
that is expected as the inflaton field $\phi$ rolls down its effective potential. On the other
hand, Planck strengthened the upper limit on the ratio tensor to scalar perturbations:$ r < 0.10$,
which restricts the energy density during inflation and poses a challenge for simple
inflationary models with potentials that are monomial powers of $\phi$. As can be seen
in Fig.~\ref{fig:inflation}, one model that does survive Planck is the original Starobinsky model
with a non-minimal $R^2$ term~\cite{Starobinsky}. Another model that makes very similar predictions is
Higgs inflation~\cite{BS}, but this would require the Higgs mass to lie in the region of electroweak
vacuum stability, which is now disfavoured, as discussed above.

An attractive alternative possibility is provided (yet again) by supersymmetry. Any
successful inflationary model requires either an inflaton mass that is $\ll M_P$ and/or
an inflaton self-coupling that is $\ll$ unity, either of which would be technically natural
in a supersymmetric model. Remarkably, the simplest single-field,
globally-supersymmetric Wess-Zumino model has an effective potential that can inflate 
successfully in agreement with Planck data~\cite{CEM},
as seen in Fig.~\ref{fig:inflation}. Even more remarkably, a locally-supersymmetric
version in the framework of no-scale supergravity yields the same effective potential as
the Starobinsky model~\cite{Cecotti,ENO6}. Maybe, as we wrote a while back, inflation cries out for
supersymmetry~\cite{ENOT}!

\section{Dark Matter}

It does not need repeating here that supersymmetry provides a very natural
candidate for cold dark matter, and that experiments are hot on its trail~\cite{RO}.
Many experiments are searching directly for dark matter scattering via either
spin-independent and/or Ðdependent interactions. XENON100 is currently
the market leader in direct searches for spin-independent dark matter interactions
over a large range of masses. There is some confusion at low masses, with
positive claims from DAMA/LIBRA, CoGENT, CRESST and now CDMS that are
difficult to reconcile with the limit from XENON100. The low-mass region can be
explored in searches for mono-photon, -$Z$ and -$W$ events at the LHC, but
the interpretations of these searches are somewhat model-dependent, since
they are sensitive to possible form factors in the interactions between dark
matter and quarks or gluons~\cite{BDM}. Future direct dark matter search experiments such as
XENON1T, DARWIN and EURECA will reach deeply into the parameter spaces of
simple supersymmetric models.

Indirect searches for astrophysical dark matter particles via their annihilation
products have generated some excitement recently~\cite{RO}.  For example, the AMS-02
experiment has recently confirmed~\cite{AMS} the rising positron fraction observed by
previous experiments. However, explaining this via dark matter annihilations
would require a very large boost factor, and is subject to important restrictions
from the absence of signals in $\gamma$ rays and antiprotons~\cite{LB}. More excitement
was generated by the report of a possible $\gamma$-ray line at $\sim 130$~GeV
in the FERMI satellite data~\cite{LB}. However, this would also require a very large
annihilation cross section, and any fundamental explanation must deal with
the fact that the line is also observed in $\gamma$ rays from the Earth's limb!
This line sounds to me like a systematic effect.

\section{Supersymmetry}

In addition to all the traditional arguments for supersymmetry based on
naturalness, unification, string, dark matter, etc., the LHC discovery of a
Higgs boson has provided more. Not only could supersymmetry stabilize
the electroweak vacuum, but it predicted successfully that the Higgs mass
should be $< 130$~GeV in simple models~\cite{ERZ}, and it also predicted successfully that its couplings
should be within a few \% of the Standard Model values~\cite{MC8}.

\begin{figure*}[htb!]
\begin{centering}
\hspace{3cm}
\includegraphics[scale=0.4]{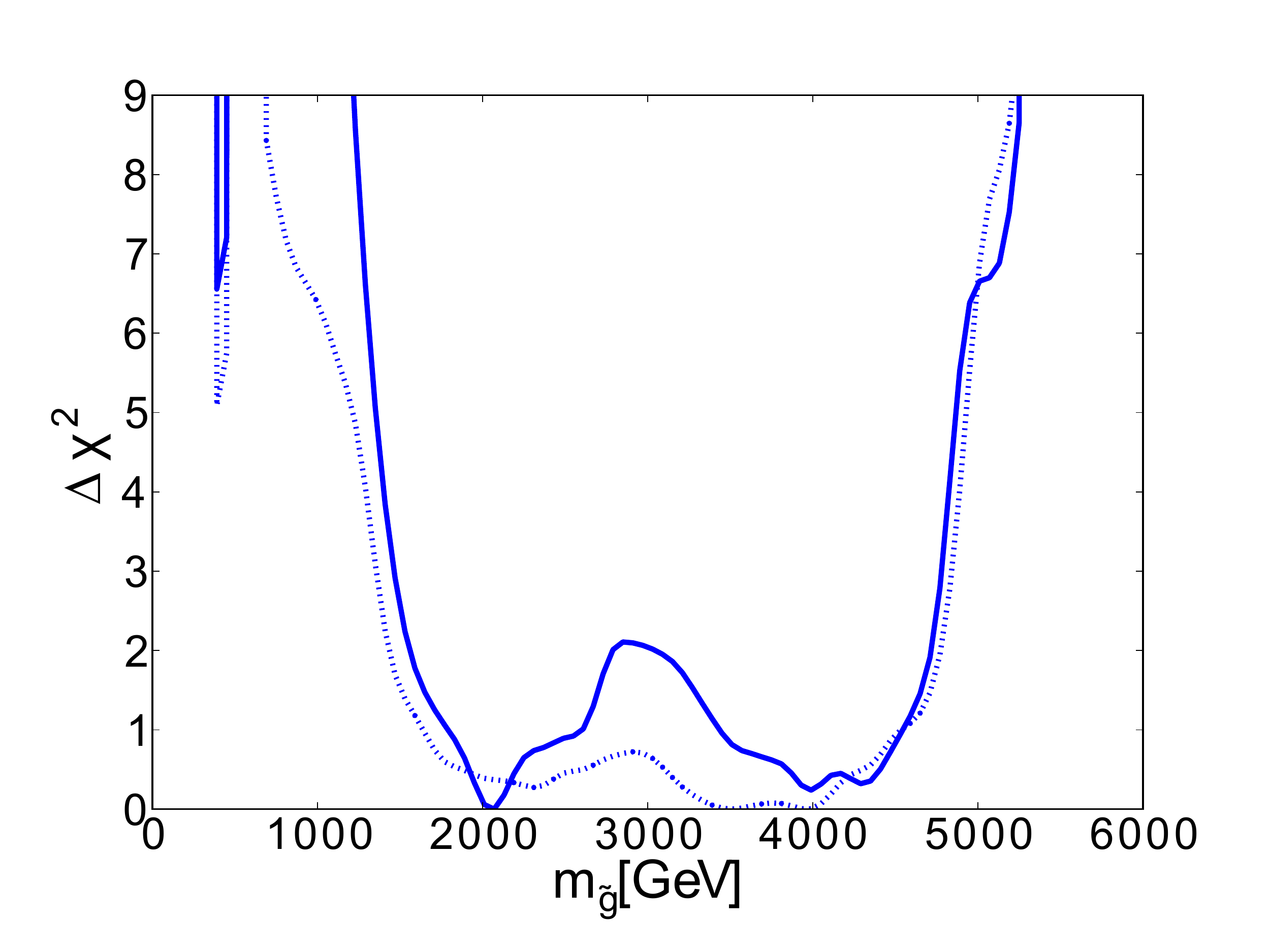}
\end{centering}
\caption{\it The $\chi^2$ likelihood in the (oversimplified) constrained minimal
supersymmetric extension of the Standard Model (CMSSM)
as a function of  the gluino mass, based on a global fit to
the LHC with 5/fb at 8~TeV and XENON100 data set (solid lines), compared to a fit to
the LHC data from 1/fb at 7~TeV (dashed  lines)~\cite{MC8}.
}
\label{fig:MC8}
\end{figure*}

Global fits to supersymmetric model parameters incorporate inputs
from precision electroweak observables, flavour physics, the anomalous magnetic
moment of the muon ($g_\mu - 2$), the Higgs mass, the dark matter density
and searches for astrophysical dark matter, and LHC constraints from searches
for missing-energy (MET) events~\cite{AP} and heavy Higgs bosons~\cite{CM}. The Higgs mass and
MET searches push strongly-interacting sparticle masses above 1~TeV, which makes
$g_\mu - 2$ difficult to explain within simple models. Fig.~\ref{fig:MC8} displays
the $\chi^2$ function for the gluino mass found in the simplest supersymmetric
model with universal soft supersymmetry breaking~\cite{MC8}. Gluino masses below about
3~TeV should be accessible to the high-luminosity upgrade of the LHC (HL-LHC),
but gluino masses above that would probably be accessible only to a higher-energy
collider.

Many such analyses have been made under simplifying assumptions, and one
might wonder how general their conclusions may be. A full exploration of the
multi-dimensional supersymmetric parameter space is beyond our computational reach.
However, one promising approach may be to reduce its effective dimensionality by
demonstrating that the limits from some combination of LHC searches is independent
of many aspects of the sparticle spectrum~\cite{BMa}. A suitable extension of this
approach might offer some reassurance that we have missed some low-mass scenario
between the supersymmetric lamp-posts where we have searched.

My point of view is that supersymmetry anywhere below the grand unification or Planck scale
is better than nowhere, and so should be part of the big picture. Moreover,
we have always understood that supersymmetry could not explain the hierarchy,
merely help stabilize it, so the need for new ideas is not new!

\section{Future Accelerators}

In view of the Higgs boson discovery, it is natural to consider building a `Higgs Factory',
capable of producing many Higgs bosons and measuring its properties in great detail~\cite{RA}.
The good news is that we already have a Higgs Factory - it is called the LHC, which
has already produced millions of Higgs bosons, though only a fraction have been
observed. However, the HL-LHC would have impressive capabilities for Higgs
measurements~\cite{HM}. Fewer Higgs bosons but much cleaner experimental conditions would be 
provided by an $e^+ e^-$ collider, which might be either linear (ILC~\cite{ILC}, CLIC~\cite{CLIC}) or circular 
(LEP3, TLEP~\cite{TLEP}). Other possibilities include a photon-photon or muon collider. An ICFA
workshop at Fermilab in November 2012 considered these various possibilities, and
concluded that circular $e^+ e^-$ colliders offered prospects for the highest precision in
Higgs studies~\cite{ICFA}.

In considering options for the future, we should be mindful that the HL-LHC is by no
means a `done deal': it needs needs global collaborative R\&D on a number of high-tech
items such as high-field dipoles and quadrupoles, long high-temperature
superconducting links, crab cavities, etc., as well as substantial detector upgrades~\cite{AC}. 
Subsequent priorities, whether an $e^+ e^-$ 
collider, a higher-energy hadron collider in the LHC tunnel, or a large circular tunnel
capable of housing an $e^+ e^-$ collider or a very-high-energy hadron collider,
will depend on what the LHC experiments find when its energy is increased in 2015
to 13 or 14~TeV, and no decision should be made before its results are available.

In this context, it is worth recalling some key phrases from the recently-agreed
European Strategy for Particle Physics~\cite{MK}, which states that ``Europe's top priority should 
be the exploitation of the full potential of the LHC, including the high-luminosity upgrade ".
It goes on to say that ``CERN should undertake design studies for accelerator projects in a 
global context, with emphasis on proton-proton and electron- positron high-energy frontier machines."
Then it says that ``The initiative from the Japanese particle physics community to host the ILC in 
Japan~\cite{HM} is most welcome, and ... Europe looks forward to a proposal from Japan to discuss a 
possible participation."

We have been reminded here that big accelerator laboratories have important
r{\^ o}les in innovation, training and outreach as well as research~\cite{RH}. They
push forward the frontiers of technology as well as science, they can
stimulate young people to study STEM subjects, and help society
to appreciate science. A successful accelerator project requires a
sustained commitment and global collaboration among many partners,
in pursuit of common goals. The health of the field requires major accelerator 
projects in all regions of the world, but it is important not to underestimate the
complex scientific, technical, administrative and political challenges involved.
It is important to learn the lessons from the last major `green-field' project in particle physics,
and propose and discuss any major future project in an international context
from the beginning.

Paraphrasing a conversation I had with Mrs. Thatcher
when we were introduced at CERN in 1982, I said that ``My job as a theorist
is to think of thing for the experiments to look for, but then we hope they find
something different." Mrs. Thatcher liked things to be the way she wanted them to be, so
she asked ``Wouldn't it be better if they found what you predicted". I responded along the lines
that ``If they just find exactly what we predict, we would have no clues how to progress."
In this spirt, let us all hope that the Higgs boson is not exactly that of the
Standard Model, and that higher-energy LHC running will reveal other new
physics beyond the Standard Model.

\section*{Acknowledgments}
This work is supported partly by the London
Centre for Terauniverse Studies (LCTS), using funding from the European
Research Council 
via the Advanced Investigator Grant 267352. I thank the CERN
Theory Division for its kind hospitality.

\end{document}